\documentclass[letter]{aa}  

\usepackage{graphicx}
\usepackage{txfonts}
\begin{document} 

\title{Possible pair-instability supernovae at solar metallicity from magnetic stellar progenitors}

   \author{Cyril Georgy
          \inst{1}
          \and
          Georges Meynet\inst{1}
          \and
          Sylvia Ekstr\"om\inst{1}
          \and
          Gregg A. Wade\inst{2}
          \and
          V\'eronique Petit\inst{3}
          \and
          Zsolt Keszthelyi\inst{2,4}
          \and
          Raphael Hirschi\inst{5,6}
          }

   \institute{Geneva Observatory, University of Geneva, Maillettes 51, 1290 Sauverny, Switzerland\\
              \email{cyril.georgy@unige.ch}
         \and
             Department of Physics, Royal Military College of Canada, PO Box 17000 Station Forces, Kingston, ON, Canada K7K 0C6
         \and
             Department of Physics and Space Sciences, Florida Institute of Technology, 150 W. University Blvd, Melbourne, FL 32904
         \and
            Department of Physics, Engineering Physics and Astronomy, Queen's University, 99 University Avenue, Kingston, ON K7L 3N6, Canada
         \and
             Astrophysics group, Keele University, Keele, Staffordshire ST5 5BG, UK
         \and
             Kavli IPMU (WPI), Tokyo Institutes for Advanced Study, The University of Tokyo, 5-1-5 Kashiwanoha, Chiba 277-8583, Japan
             }

   \date{}

 
  \abstract
   {Near-solar metallicity (and low-redshift) Pair-Instability Supernova (PISN) candidates challenge stellar evolution models. Indeed, at such a metallicity, even an initially very massive star generally loses so much mass by stellar winds that it will avoid the electron-positron pair-creation instability. We use recent results showing that a magnetic field at the surface of a massive star can significantly reduce its effective mass-loss rate to compute magnetic models of very massive stars (VMSs) at solar metallicity and explore the possibility that such stars end as PISNe. We implement the quenching of the mass loss produced by a surface dipolar magnetic field into the Geneva stellar evolution code and compute new stellar models with an initial mass of $200\,M_\sun$ at solar metallicity, with and without rotation. It considerably reduces the total amount of mass lost by the star during its life. For the non-rotating model, the total (CO-core) mass of the models is $72.8\,M_\sun$ ($70.1\,M_\sun$) at the onset of the electron-positron pair-creation instability. For the rotating model, we obtain $65.6\,M_\sun$ ($62.4\,M_\sun$). In both cases, a significant fraction of the internal mass lies in the region where pair instability occurs in the $\log(T)-\log(\rho)$ plane. The interaction of the reduced mass loss with the magnetic field efficiently brakes the surface of the rotating model, producing a strong shear and hence a very efficient mixing that makes the star evolve nearly homogeneously. The core characteristics of our models indicate that solar metallicity models of magnetic VMSs may evolve to PISNe (and pulsation PISNe).}


   \maketitle
%

\section{Introduction}

The discovery of pair instability supernovae (PISNe) candidates\footnote{\footnotesize{Alternative explanations for these objects exist, see e.g. \citet{Moriya2010a}.}} at various redshifts in the last decade \citep[e.g. \citealt{Cooke2012a}, and including relatively low redshift one, thus with a relatively high metallicity,][]{GalYam2009a} has triggered great interest in understanding their formative pathways, and initial attempts to model the evolution of the progenitors of such explosions. For a PISN explosion to occur, a significant fraction of the stellar mass must be located in the range of temperatures and densities where the pair-production instability occurs, leading to an adiabatic exponent $\Gamma_\text{1, ad} < 4/3$, which has a destabilising effect\footnote{\footnotesize{$\Gamma_\text{1, ad}$ is defined as $\Gamma_\text{1, ad} = \left.\frac{\partial\left(\ln P\right)}{\partial\left(\ln\rho\right)}\right|_\text{ad}$.}} \citep[see e.g.][]{Maeder2009a}. This situation is reached for models with a CO-core mass in the range of about $60$ to $130\,M_\sun$ \citep{Heger2002a,Chatzopoulos2012a}.

A couple of superluminous supernovae (SLSNe) have been observed at solar and even higher metallicities \citep{Lunnan2014a}. These supernovae might be associated to PISNe. With the high mass-loss rates typical of hot massive stars at solar metallicity, it is very unlikely for a very massive star (VMS) to match the required end-of-life conditions to produce a PISN \citep{Yusof2013a}. As a consequence, the existence of PISNe at such a metallicity would be a challenge for our understanding of the evolution of VMSs.

However, about $7\%$ of Galactic O-type stars exhibit a measurable surface magnetic field \citep{Wade2014a}\footnote{\footnotesize{This means a surface magnetic field larger than about $100\,\text{G}$.}}. It has been shown that a surface dipolar magnetic field of sufficient strength (typically larger than about $1\,\text{kG}$) can considerably affect the winds from a hot massive star, by confining part of the wind and preventing it from escaping the star \citep{udDoula2008a,Bard2016a,Petit2017a}. \citet{Petit2017a} explore the effects of the quenching of the wind by a surface magnetic field for stellar models between $40$ and $80\,M_\sun$ at solar metallicity, and they show that the total mass lost during the main sequence (MS) can be considerably decreased compared to non-magnetic models.

In this paper, we apply the same procedure in the framework of VMSs at solar metallicity, to assess the possibility of obtaining suitable PISN progenitors at higher metallicity than is usually thought. In Sect.~\ref{Computation}, we summarise the physical modelling and parameters used for the stellar evolution computations. In Sect.~\ref{Results}, we discuss the impact of the inclusion of the mass-loss quenching by magnetic field on the evolution of solar metallicity VMSs, and compare our results with previous works. Finally, we present our conclusions in Sect.~\ref{Conclusions}.

\section{Physical modelling and parameters}\label{Computation}

We use the Geneva Stellar Evolution Code \citep{Eggenberger2008a} to perform the simulations of the current work, in the same configuration as in \citet{Ekstrom2012a}. We refer the reader to this work for the details of the various implementations used. Two important points have to be emphasised here:
\begin{itemize}
\item Our equation of state consists of a mixture of an ideal gas and radiation and accounts for the effects of partial ionisation. It does not include the effect of creation of electron-positron pairs, that would be necessary to perform an accurate simulation of the collapse towards a PISN \citep[e.g.][]{Chatzopoulos2015a}. However, our equation of state is suitable to provide the CO core mass and to estimate the fraction of the internal mass where pair creation occurs \citep[see the discussion of][]{Yusof2013a}.
\item We use the same scheme for the computation of the mass-loss rates as used by \citet{Ekstrom2012a}: as long as the model has a surface hydrogen mass fraction $X_\text{S}>0.3$, we use the \citet{Vink2000a,Vink2001a} rates. When the model enters the Wolf-Rayet star (WR) regime, i.e. when $X_\text{S}\leq0.3$ with $\log(T_\text{eff})>4.0$, we use either the \citet{Grafener2008a} prescription in its validity domain, or the \citet{Nugis2000a} prescription elsewhere. Note that in case the \citet{Vink2000a,Vink2001a} mass-loss rates are higher than the \citet{Grafener2008a} or \citet{Nugis2000a} rates, the former prescription is used.
\end{itemize}

We implemented mass-loss quenching by the fossil surface magnetic field in the same way as \citet{Petit2017a}, by following the time evolution of a fossil surface dipolar field\footnote{\footnotesize{In \citet{Petit2017a} the magnetic field is the (polar) dipolar field strength. In this paper, $B_\text{eq}$ is to be understood as the equatorial magnetic field \citep{udDoula2008a,Meynet2011a}, the value of which is half of the dipolar field strength.}} (by assuming flux conservation). The escaping wind fraction $f_B$ is given by:
\begin{equation}
f_B = \frac{\dot{M}}{\dot{M}_{B=0}} = 1-\sqrt{1-\frac{r_\star}{r_\text{c}}},\label{eq_fraction}
\end{equation}
where $r_c$ is the radius of the farthest closed loop of the magnetic field, and is computed as a function of the Alfv\'en radius and the confinement parameter, and $r_\star$ is the stellar radius \citep[for details, see][]{Petit2017a,udDoula2002a}.

As mentioned by \citet{Petit2017a}, in the case of a rotating star, $r_\text{c}$ should be replaced by the Keplerian co-rotation radius $r_\text{K}$ if it is smaller than $r_\text{c}$. In practice, this never occurred in our simulations.
For the computation of our rotating models, we assumed that the surface magnetic field of the star produces a braking of the surface. We imposed a torque as a boundary condition for the transport of angular momentum \citep{udDoula2008a}, as in \citet{Meynet2011a}:
\begin{equation}
\frac{\text{d}J}{\text{d}t} \approx \frac{2}{3}\dot{M}_\text{B=0}\Omega r_\star^2\left(0.3+\left(\eta_\star + 0.25\right)^\frac{1}{4}\right)^2.\label{Eq_Torque}
\end{equation}

For the purpose of this paper, we computed 2 models of $200\,M_\sun$ at solar metallicity ($Z_\sun=0.014$), with an initial surface equatorial magnetic field $B_\text{eq, ZAMS} = 1000\,\text{G}$. We assume in this work that the surface magnetic field evolves by conserving the magnetic flux throughout the whole evolution. One model is computed without rotation, and the other one with an initial equatorial velocity $V_\text{ini}/V_\text{crit} = 0.4$, where the critical velocity $V_\text{crit} = \sqrt{\frac{2GM}{3R_\text{p,b}}}$ with $R_\text{p,b}$ being the polar radius at the break-up velocity \citep{Maeder2000a}. Both models are evolved up to central oxygen burning.

\section{Results}\label{Results}

\begin{figure*}
\centering
\includegraphics[width=.34\textwidth]{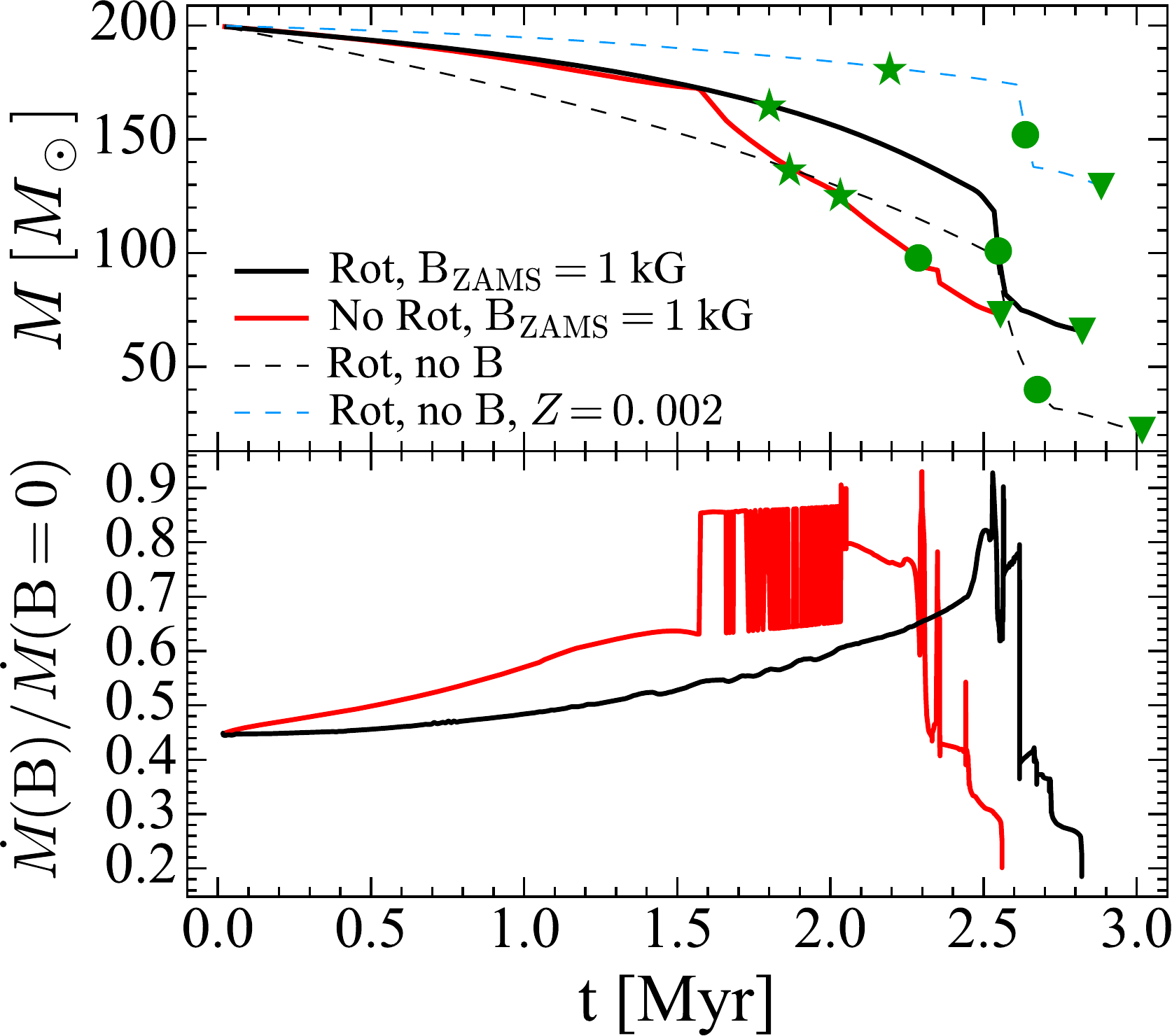}\hfill\includegraphics[width=.31\textwidth]{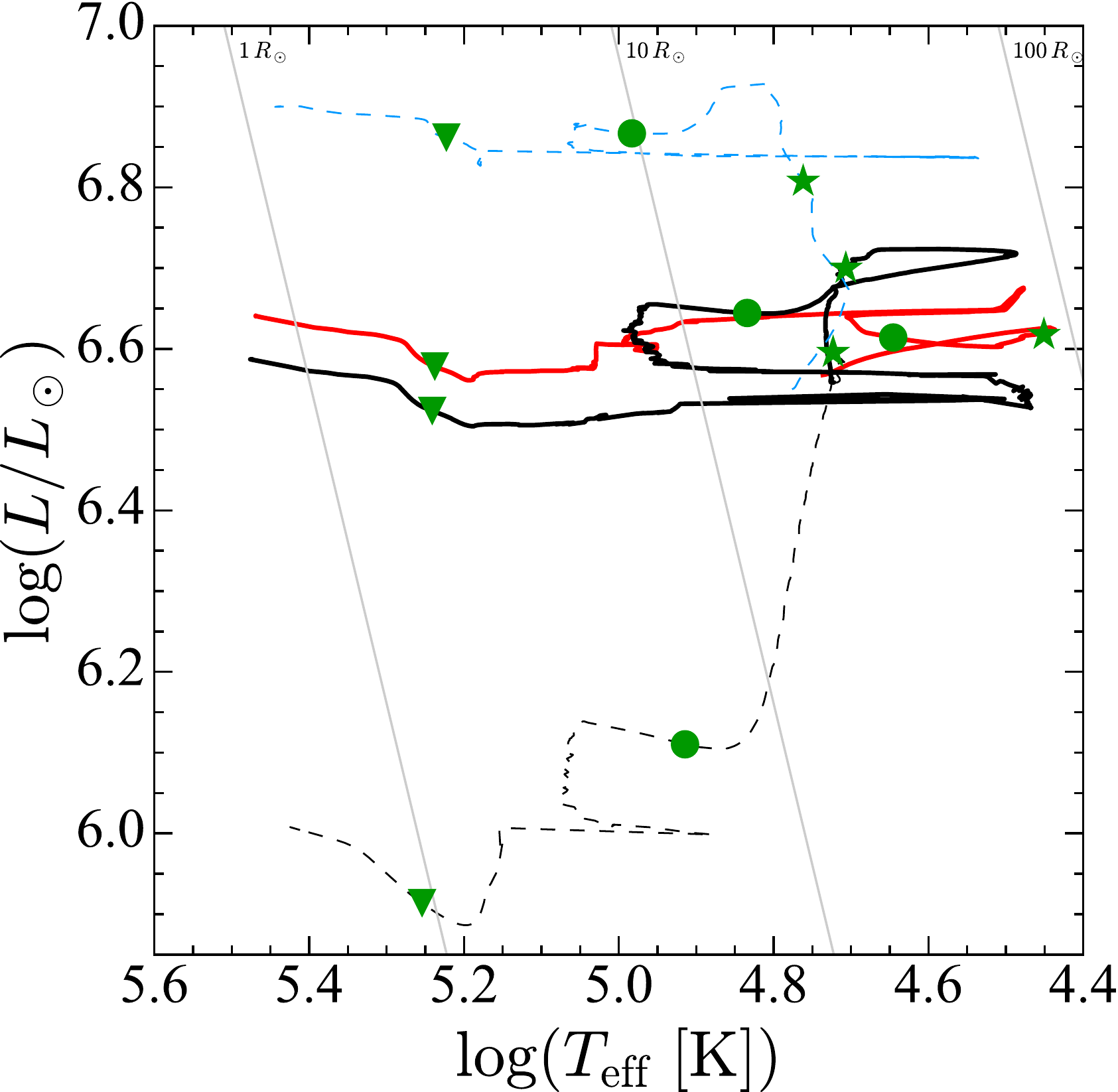}\hfill\includegraphics[width=.31\textwidth]{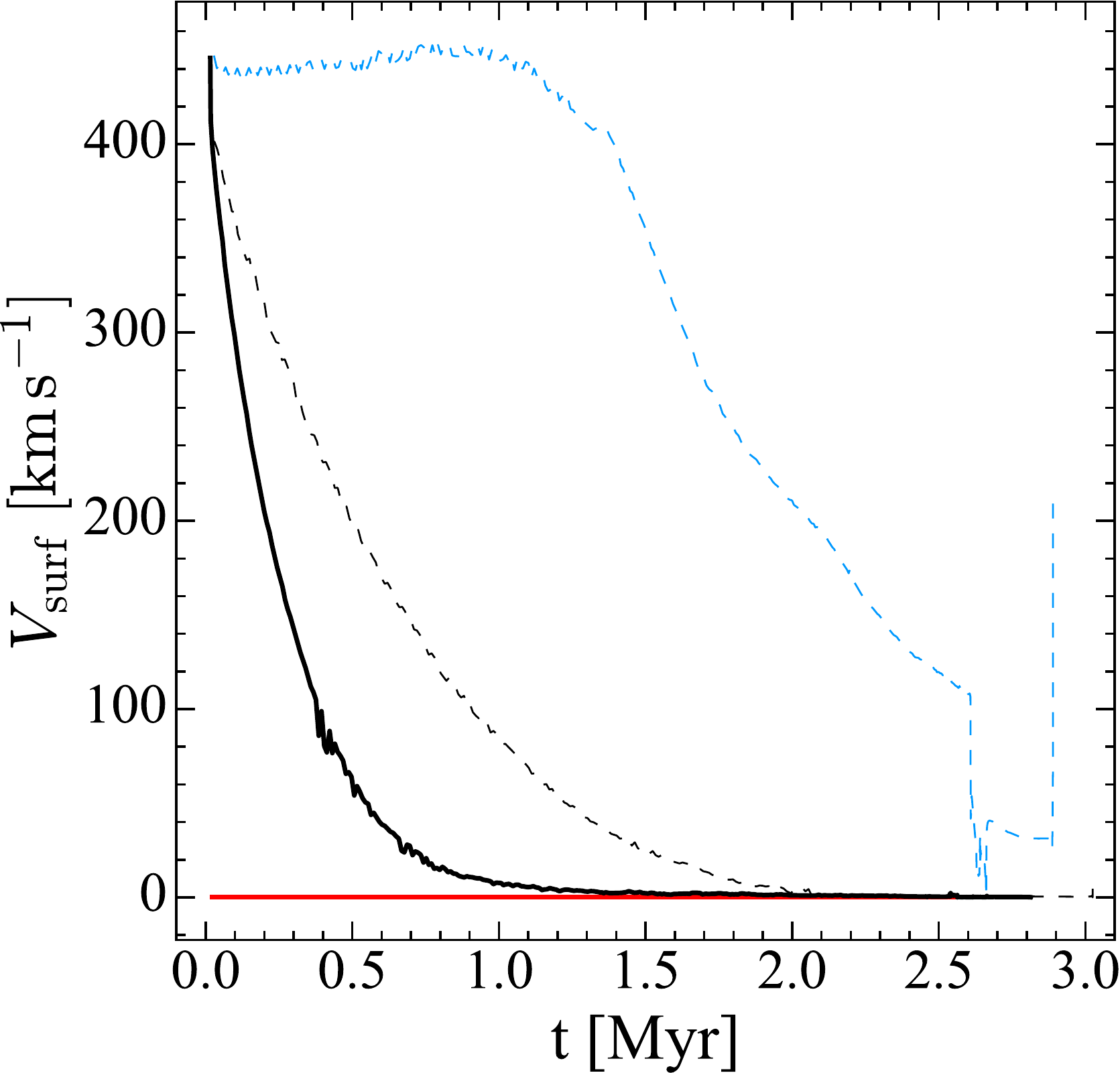}
\caption{\textit{Left panel, top:} time evolution of the total mass of the models. The magnetic models computed for this work are in black (rotating model) and red (non-rotating one). Models from \citet{Yusof2013a} are shown for comparison: non-magnetic rotating model at solar metallicity (thin dashed black line) and SMC metallicity (thin dashed blue line). Key evolutionary phases are indicated: beginning of the WR phase (star), end of the MS (circle), and end of central He burning (triangle). \textit{Left panel, bottom:} time evolution of the escaping wind fraction $f_B$ (see eq.~\ref{eq_fraction}) for the magnetic rotating (black) and non-rotating (red) models. \textit{Centre panel:} HRD tracks for the same models as in left panel. The meaning of the colours and symbols is the same. Iso-radii are indicated in light grey ($1$, $10$, and $100\,R_\sun$ from left to right). \textit{Right panel:} Time evolution of the surface velocity of our models. We use the same colour scheme as for the top-left panel.}
\label{Fig_HRD_Mass}
\end{figure*}

In this section, we discuss the most important results of our computations. We refer to our new models as ``magnetic models''. We compare our models with the following ones from \citet{Yusof2013a}: the $200\,M_\sun$ at solar metallicity, and the $200\,M_\sun$ at a lower metallicity ($Z=0.002$), corresponding to the metallicity of the Small Magellanic Cloud (SMC). We chose these models because 1) they were computed with the same stellar evolution code, so the differences are only due to the inclusion of mass-loss quenching and surface braking due to magnetic field (see previous Section); 2) the lower metallicity model has lower mass-loss rates compared to the solar metallicity one, so it is an interesting point of comparison with our new magnetic models; and 3) the SMC metallicity model retains enough mass ($129.2\,M_\sun$) to end as a PISN, whereas the rotating solar metallicity model with no magnetic quenching loses too much mass to end as a PISN 
\citep[or even as a pulsation pair instability SN, PPISN, with a final mass of $21.9\,M_\sun$, see][]{Woosley2016a}.

\subsection{Mass loss and escaping wind fraction}

The left panel of Fig.~\ref{Fig_HRD_Mass} shows the variation of the total mass (top) and of the escaping wind fraction $f_B$ defined in Eq.~\ref{eq_fraction} for the magnetic models (thick lines, rotating model in black and non-rotating one in red), and for the solar-metallicity model (black thin dashed line) and SMC-metallicity one (blue thin dashed line) from \citet{Yusof2013a}. During the MS, $\eta_\star$ decreases due to the evolution of the stellar parameters, making the escaping wind fraction grow from a value around $0.5$ on the ZAMS to $0.8$ near the end of the MS. For the non-rotating model, the oscillations between about $1.5$ and $2\,\text{Myr}$ are due to successive crossings of the bistability limit \citep{Pauldrach1990a,Vink2000a,Vink2001a}. The mass-loss rate on the cool side is considerably higher than on the hot side, therefore the escaping wind fraction is higher on the cool side. The change in the mass-loss rate keeps the star at the bi-stability limit for a while, generating the oscillations. After the MS, both models become very hot and compact Wolf-Rayet (WR) stars (see the iso-radii in the right panel of Fig.~\ref{Fig_HRD_Mass} and Sect.~\ref{HRDTracks}), with a very low escaping wind fraction.

Magnetic wind quenching considerably reduces the total amount of mass lost during the evolution. Both magnetised models have a total mass of around $70\,M_\sun$ at the onset of central oxygen burning, while the corresponding non-magnetised model ends with about $20\,M_\sun$. Overall, our solar metallicity models with magnetic fields show an evolution of the total mass in between the solar- and SMC-metallicity non-magnetic models \citep[see also Fig.~2 of][]{Petit2017a}.

\subsection{Hertzsprung-Russell diagram tracks}\label{HRDTracks}

The middle panel of Fig.~\ref{Fig_HRD_Mass} shows the tracks in the Hertzsprung-Russell diagram (HRD) of the four models (same colour code as left panel). Due to the reduced mass-loss rates, the tracks of the magnetic models are very different from the corresponding solar-metallicity non-magnetic model: instead of evolving with decreasing luminosity, the magnetic models experience an increase in the luminosity during the MS, as is seen in lower mass or lower metallicity models. Once the mass loss has uncovered the hydrogen-burning core, the star enters a WR phase, and becomes very hot. Note that rotating models evolve in a quasi-chemically homogeneous way, hence the almost vertical tracks during the MS \citep[see e.g.][]{Maeder1987a,Yoon2005a}. The mass loss is however larger than that of the non-magnetic SMC metallicity model, causing the tracks to remain in a more limited range of luminosity than the SMC model from \citet{Yusof2013a}.

\subsection{Surface and internal velocities}

The right panel of Fig.~\ref{Fig_HRD_Mass} shows the time evolution of the surface velocity of the models. Even with a considerably reduced mass loss compared to the non-magnetic case, the surface of the rotating magnetic model experiences a much stronger braking, due to the coupling between the stellar wind and the magnetic field \citep{Meynet2011a}. The surface velocity drops below $100\,\text{km\,s}^{-1}$ in about $0.5\,\text{Myr}$. Thus, we expect VMSs with an external magnetic field to be slow rotators, even if their mass-loss rates are decreased with respect to the non-magnetic case.

In case the magnetised models avoid exploding as PISNe (see Section~\ref{Sec_PISN}), it is interesting to determine the amount of angular momentum retained in the core near the end of the evolution. As illustrated by the evolution of the surface velocities, the angular momentum lost by the magnetic models is higher than in the non-magnetic case, even though the mass-loss rate is reduced. This is due to the increased torque exerted on the surface due to the magnetic field (see Eq.~\ref{Eq_Torque}). The extraction of angular momentum from the core is efficient enough to prevent  the formation of a long-soft gamma-ray burst through the collapsar scenario \citep[see also the discussion of][]{Yusof2013a}.

\subsection{Central conditions and possible PISN progenitor}\label{Sec_PISN}

\begin{figure*}
\centering
\includegraphics[width=.32\textwidth]{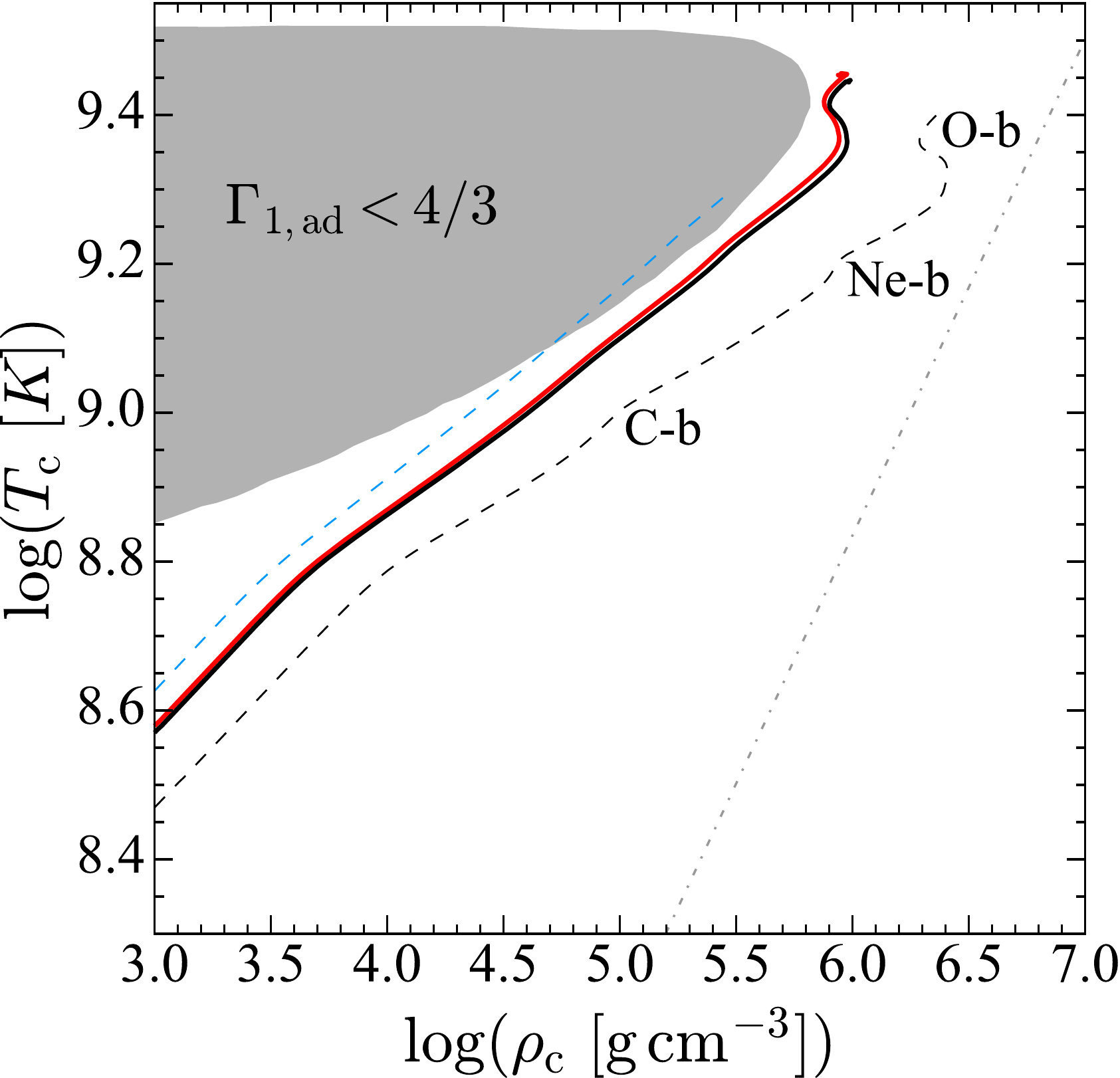}\hfill\includegraphics[width=.32\textwidth]{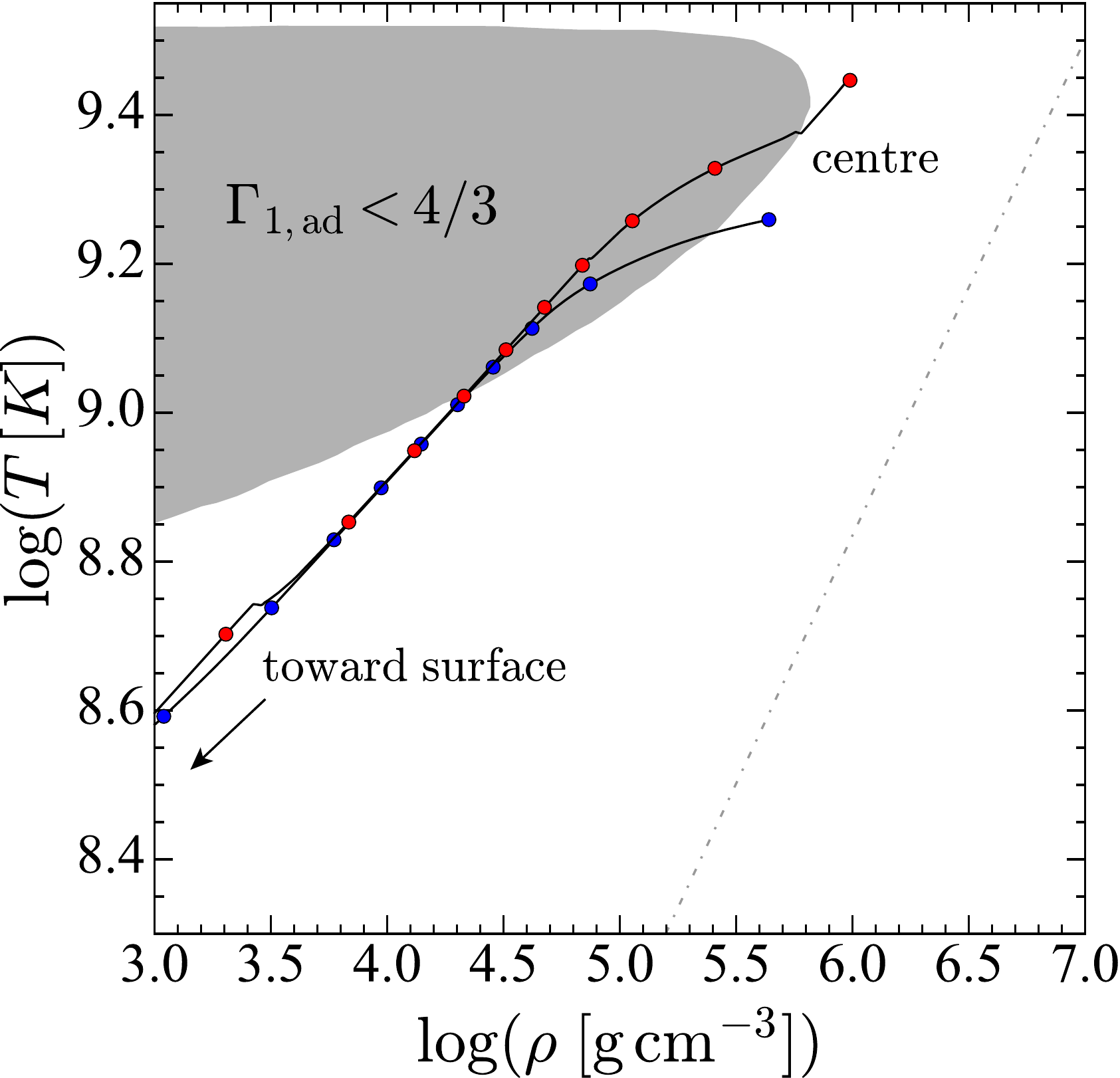}\hfill\includegraphics[width=.31\textwidth]{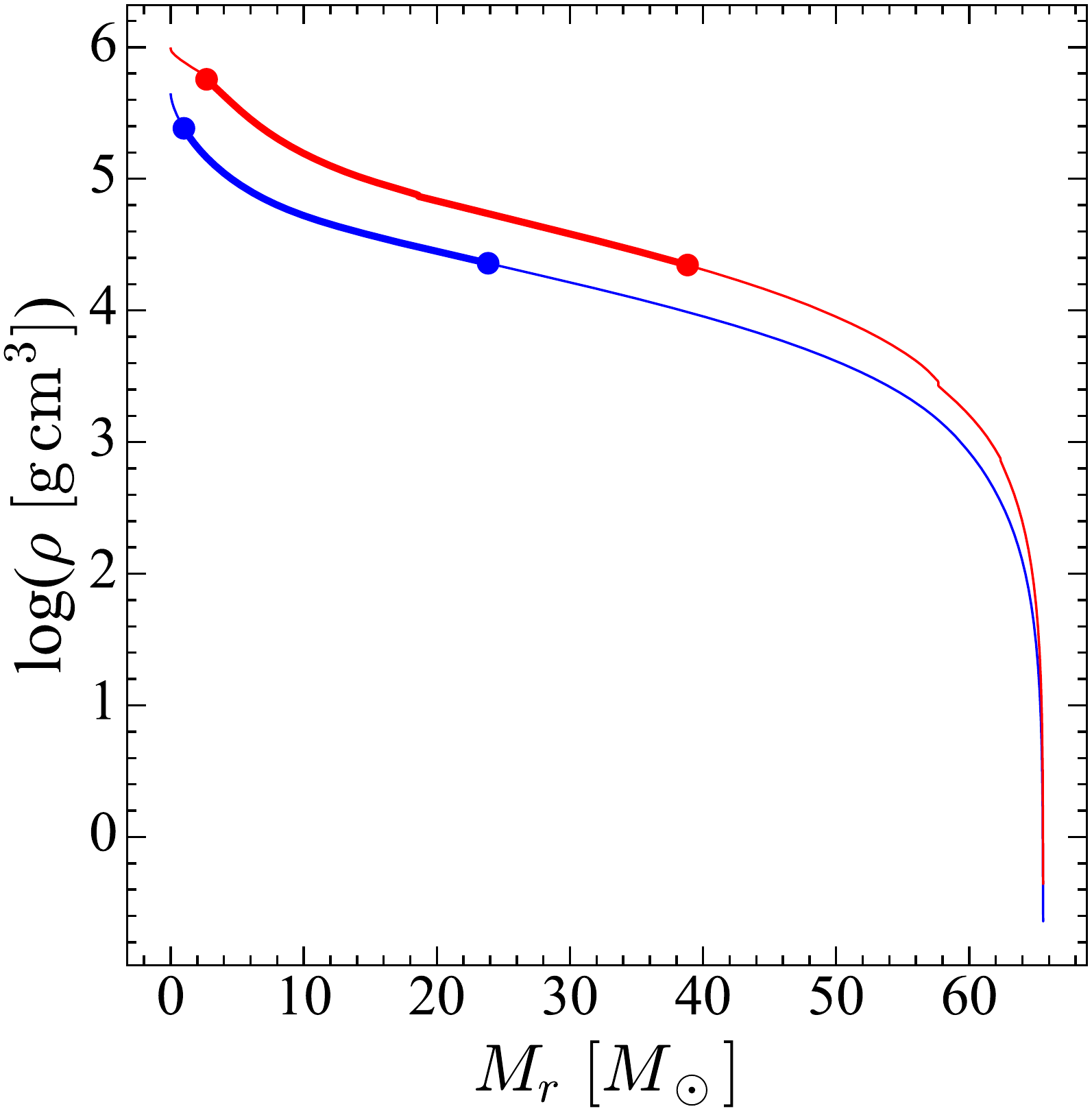}
\caption{\textit{Left panel:} Evolutionary tracks in the central temperature versus central density diagram. The colours of the tracks have the same meaning as in top-left panel of Fig.~\ref{Fig_HRD_Mass}. The thin grey dotted-dashed line represents the transition between the perfect gas and degenerate gas domains. The shaded grey zone shows the region of the diagram where electron-pair creation occurs, from \citet{Chatzopoulos2015a}. \textit{Centre panel:} Internal profiles of the magnetic rotating model at the beginning (blue dots) and end (red dots) of central oxygen burning. The dots are spaced by $10\%$ of the total mass of the star. \textit{Right panel:} Internal density profiles (as a function of the mass coordinate) at the beginning (blue) and end (red) of central oxygen burning. The thick part of the curves indicate the region inside the star inside the electron-pair creation region.}
\label{Fig_Tcrhoc}
\end{figure*}

The evolution of the central temperature as a function of the central density is shown in the left panel of Fig.~\ref{Fig_Tcrhoc}. As expected from the time evolution of the total mass (top-left panel of Fig.~\ref{Fig_HRD_Mass}), both non-rotating and rotating magnetic models evolve in between the solar-metallicity and SMC-metallicity model from \citet{Yusof2013a}. Contrary to the SMC-metallicity model, which enters the temperature-density domain where electron-pair creation occurs \citep[grey shaded area, from][]{Chatzopoulos2015a}, both our magnetic models only graze this region. However, a significant fraction of the internal mass of these models crosses it (centre panel). This is illustrated by the thin black lines with blue (beginning of central oxygen burning) or red (end of central oxygen burning) dots, showing the temperature-density profile inside the magnetic rotating model. The dots are positioned every $10\%$ of the internal mass. At the beginning of the central oxygen burning, about $30\%$ of the internal mass is in the region with $\Gamma_\text{1, ad} < 4/3$. This fraction reaches about $50\%$ at the end of central oxygen burning. The right panel of Fig.~\ref{Fig_Tcrhoc} shows the internal density profile as a function of the mass coordinate. The thick parts of the curves indicate the region inside the star lying inside the pair-creation region at the beginning (blue) and end (red) of central oxygen burning, illustrating the contraction of the model during this phase, as well as the growth of the region inside the pair-creation region with time.

Moreover, the mass of the CO core of our magnetic models is $70.1\,M_\sun$ (non-rotating model) and $62.4\,M_\sun$ (rotating model). It is in the range of mass leading to PISN explosion according to \citet{Heger2002a}. With the large fraction of the internal mass inside the region with $\Gamma_\text{1, ad} < 4/3$ , it is thus plausible \citep[see][]{Fowler1964a} that VMS models with magnetic quenching of the stellar winds lead to PISN progenitors at a higher metallicity than usually considered when a standard mass loss is applied. Models with the same initial mass but a slightly higher initial magnetic field, or models with a slightly higher initial mass, will also enter more deeply inside the $\Gamma_\text{1, ad} < 4/3$ region. 

We emphasise also that our choice of initial magnetic field strength is modest, and hence the mass-loss quenching we obtain is small compared to what it would be with a higher initial field. Moreover, our models always evolve at a high enough effective temperature to never have a large external convective zone that would probably destroy the fossil magnetic field hypothesised in this work. Another point is that due to the strong mass-loss, layers that were previously located inside the convective core, where a dynamo could be active, are uncovered relatively quickly. The effect on the global dipolar field is unknown, and beyond the scope of this paper. If our assumption of flux conservation breaks down at some point in the evolution, the total mass lost by our models would be larger, and hence disfavour our scenario for a high-metallicity PISN progenitor. Additional mass loss may occur as the Eddington factor reaches high values towards the end of the evolution \citep[see Fig. 9 in][and the corresponding discussion]{Yusof2013a}. In any case, magnetic quenching is more favorable to VMSs at high(er) metallicities undergoing the pulsation pair-instability \citep[see][and references therein]{Woosley2016a}.

Our modelling considers steady state mass loss. As our models evolve close to the Eddington limit ($L/L_\text{Edd} = 0.5 - 0.7$ during the MS, but up to 0.9 at the very end of the evolution), it is possible that eruptive mass-loss events may occur. However, such events are currently unpredictable and are not modelled in stellar evolution codes. It is thus difficult to quantify their impact on the scenario discussed here.

\section{Conclusions}\label{Conclusions}

Observations show that about $7\%$ of O-type stars have an important fossil surface dipolar magnetic field at solar metallicity, ranging from a few $100\,\text{G}$ to about $20\,\text{kG}$ \citep{Wade2014a}. It has been shown that such surface magnetic fields are able to significantly quench the mass loss from the star \citep{Petit2017a}.

In this paper, we explore the impact of such a mass-loss quenching on the evolution of VMSs at solar metallicity, in case of a modest surface magnetic field of $1000\,\text{G}$. We show that this mechanism is able to reduce the total mass lost during the evolution in such a way that the models end with a massive CO core and with a significant fraction of their internal mass in the density and temperature domain where electron-pair creation is possible. Such a mechanism leads thus to a plausible scenario to form PISN at higher metallicity than is usually thought, and could help in explaining SLSNe in the local Universe. The rarity of VMSs at near-solar metallicity \citep{Martins2015c} and the small fraction of massive stars exhibiting fossil surface magnetic fields provides a natural explanation of the scarcity of such events at high metallicity \citep{Lunnan2014a}.

In the future, we intend to explore more systematically the parameter space of the initial mass $M_\text{ZAMS}$ and initial magnetic field strength $B_\text{ZAMS}$, to determine the initial parameters leading to PISNe (and pulsation-PISNe). A complete study of the possibility of PISN explosion and of the lightcurve expected is also planned, as was done for models from \citet{Yusof2013a}  in \citet{Kozyreva2017a}.

\begin{acknowledgements}
CG, GM, and SE acknowledge support from the Swiss National Science Foundation (project number 200020-160119). GAW acknowledges Discovery Grant support from the Natural Sciences and Engineering Research Council (NSERC) of Canada. VP acknowledges support provided by the NASA through Chandra Award Number GO3-14017A issued by the Chandra X-ray Observatory Center, which is operated by the Smithsonian Astrophysical Observatory for and on behalf of the NASA under contract NAS8-03060. RH acknowledges support from EU-FP7-ERC-2012-St Grant 306901 and from the World Premier International Research Center Initiative (WPI Initiative, MEXT, Japan).
\end{acknowledgements}

\bibliographystyle{aa}
\bibliography{MyBiblio}
\end{document}